\documentclass[default,iicol]{sn-jnl}% Default with double column layout

\usepackage{amsmath}
\usepackage{mathtools, cuted}
\usepackage{caption}
\usepackage{subcaption}
\usepackage[english]{babel}
\usepackage{lipsum}
\usepackage{url}

\theoremstyle{thmstyleone}%
\theoremstyle{thmstyletwo}%

\theoremstyle{thmstylethree}%

\begin{document}

\title[mode = title]{CerviFormer: A Pap-smear based cervical cancer classification method using cross attention and latent transformer}
%Integrating cross attention and latent transformer for cervical cancer classification using pap-smear images} 

\author[1]{\fnm{Bhaswati} \sur{Singha Deo}}

\author*[2]{\fnm{Mayukha} \sur{Pal}}\email{mayukha.pal@in.abb.com}
\author[3]{\fnm{Prasanta} \sur{K.Panigarhi}}

\author*[1,4]{\fnm{Asima} \sur{Pradhan}}\email{asima@iitk.ac.in}

\affil[1]{\orgdiv{Center for Lasers and Photonics}, \orgname{Indian Institute of Technology Kanpur}, \city{Kanpur}, \postcode{208016}, \country{India}}

\affil*[2]{\orgdiv{ABB Ability Innovation Center}, \orgname{Asea Brown Boveri Company}, \city{Hyderabad}, \postcode{500086}, \country{India}}

\affil[3]{\orgdiv{Department of Physical Sciences}, \orgname{Indian Institute of Science Education and Research Kolkata, Mohanpur}, \city{Nadia}, \postcode{741246}, \country{India}}

\affil*[4]{\orgdiv{Department of Physics}, \orgname{Indian Institute of Technology Kanpur}, \city{Kanpur}, \postcode{208016}, \country{India}}

\abstract{\textbf{Purpose:} Cervical cancer is one of the primary causes of death in women. It should be diagnosed early and treated according to the best medical advice, as with other diseases, to ensure that its effects are as minimal as possible. Pap smear images are one of the most constructive ways for identifying this type of cancer. This study proposes a cross-attention-based Transfomer approach for the reliable classification of cervical cancer in Pap smear images. \\
 \textbf{Methods:} In this study, we propose the CerviFormer- a model that depends on the Transformers and thereby requires minimal architectural assumptions about the size of the input data. The model uses a cross-attention technique to repeatedly consolidate the input data into a compact latent Transformer module, which enables it to manage very large-scale inputs. We evaluated our model on two publicly available Pap smear datasets.\\
\textbf{Results:} For 3-state classification on the Sipakmed data, the model achieved an accuracy of 93.70\%. For 2-state classification on the Herlev data, the model achieved an accuracy of 94.57\% .\\
\textbf{Conclusion:} Experimental results on two publicly accessible datasets demonstrate that the proposed method achieves competitive results when compared to contemporary approaches. The proposed method brings forth a comprehensive classification model to detect cervical cancer in Pap smear images. This may aid medical professionals in providing better cervical cancer treatment, consequently, enhancing the overall effectiveness of the entire testing process.}




\maketitle

\section{Introduction}{\!}

\begin{figure*}
	\centering  \includegraphics[width=\textwidth,height=3.2in]{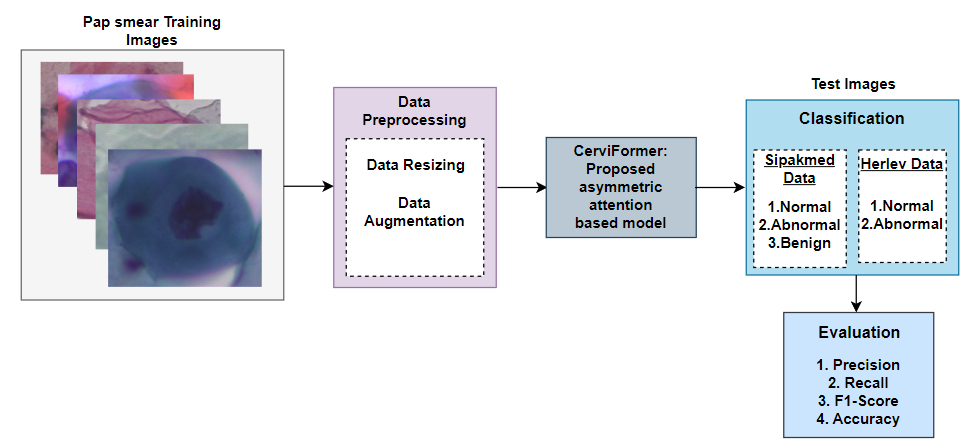}
	\caption{Overview of using CerviFormer, the proposed methodology for cervical cancer image classification}
	\label{Figure 1}
\end{figure*}

Cervical cancer is a form of cancer that grows from the cervix. It results from uncontrolled cell proliferation that can spread to other body parts. Owing to inadequate hygienic practices and availability of medical facilities, cervical cancer mortality is significantly greater in low-income and developing countries \citep{34}. The most common cause of cervical cancer is human papillomavirus (HPV) \citep{schiffman2007human}. The Pap smear test is a cervical screening procedure used to find potentially malignant conditions. The Pap test takes cervix cells and examines them for dysplasia or precancerous changes under a microscope. Abnormal findings are frequently followed by more precise diagnostic techniques, to prevent the growth of the cancer. The two essential characteristics of a cervical cell are the cytoplasm and nucleus. The cytoplasm is generally much larger than the nucleus in healthy cells. Dysplastic cells have a higher ratio of nuclei to the cytoplasm. Pap smear examinations are performed under the microscope by cytotechnicians. This test has helped to save the lives of millions of women. Pap tests are effective, although the analysis is not always accurate \cite{priest2007pathways}. Automating cell assessment might improve accuracy. \\
 A variety of nonlinear dynamics and machine learning model are being used for classification of biological signals and for predictive purposes \cite{pal2022coupled,pal2022multi,pal2021unstable}.
 %One of the areas of interest for researchers among the different Deep Learning applications in Biomedical Engineering is its use in biomedical signal processing to extract, analyze, and classify various signals or images for diagnostic needs. The majority of the biological signals are generated in nonlinear, time-varying conditions \cite{pal2022coupled,pal2022multi,pal2021unstable}. Electrocardiograms (ECGs), Electroencephalograms (EEGs), Electromyograms (EMGs), and other bioelectric signals may be processed and classified using machine learning techniques.
Computer-aided detection (CAD) techniques are widely being used to examine pap smears quickly and reliably in place of manual diagnosis \cite{fekri2022cervical,hussain2020comprehensive}. A significant number of images can be processed by the computerized system, which is beneficial for clinical monitoring, tailored medication, and comparative study \cite{marinakis2009intelligent}. Deep learning methods are being widely employed in diverse domains, including  medical imaging, computer vision, natural language processing (NLP), etc \cite{lecun2015deep,deo2022ensemble,pal2021integrative,aditya2022local,dwivedi2022detection,dwivedi2022identification,deo2022supremacy,liu2022aspect}. The fundamental drawback of conventional CAD systems is the need for manual feature extraction. This does not ensure the best classification outcomes. Deep learning (DL) is a holistic method that can automatically determine features. DL methodologies outperform traditional machine learning-based methods in image analysis domain. Conversely, a large number of labeled datasets are needed to train a deep learning model \cite{landau2019artificial}. Although deep learning produces good results, a major concern is that many of its features are inexplicable to researchers \cite{lawrence2018deep}.\\
Convolutional neural networks (CNNs), a deep learning technique, is widely used in medical image analysis \cite{pramanik2022fuzzy}. CNNs have lately been outperformed by Transformers built on the architecture of the attention mechanism \cite{vaswani2017attention,dosovitskiy2020image, tay2022efficient}. 
Transformers are highly adaptable architectural structures since they make least assumptions of the received inputs, however, memory and computational complexity rise quadratically with the number of inputs. Transformers have shown remarkable outcomes in the field of image analysis.
To tackle the computational problem, Transformers utilizes a wide range of approaches such as 2D convolution to process the pixels of an image \cite{dosovitskiy2020image,touvron2021training}, factorizing the input image into rows and columns \cite{ho2019axial,child2019generating} and different subsampling methods \cite{chen2020generative}.\\
%however, this technique relies on the grid arrangement of the pixels to minimize computational complexity, utilizes a 2D convolution to process the pixels \cite{dosovitskiy2020image,touvron2021training}, performs aggressive subsampling \cite{chen2020generative}, or by factorizing the input image into rows and columns \cite{ho2019axial,child2019generating}.
In our study, we suggest a method that has the expressivity and adaptability required to manage a wide range of input configurations while still being able to handle high-dimensional inputs.
% Our key focus is the formation of attention bottleneck by introducing a finite number of latent modules across which the inputs must transit as shown in Fig.\ref{Figure2}. As a result, the quadratic scaling issue of a standard Transformer is resolved and it further permits the formation of extremely deep models by breaking the relationship between the depth of the network and the size of the input. The suggested model iteratively focuses its limited resources on the most significant inputs. However, many modalities require spatial or temporal information, and it is often necessary to identify input among various modalities in multimodal contexts. The absence of explicit structures in our framework can be compensated by creating modality and position-specific features for every input element such as pixel. This technique of tagging input units is quite comparable to the labeled lined approach used in the construction of cross-sensory and topographic maps in biological neural networks \cite{kandel2012principles}.\\
We propose an iterative attention-based method to classify cervical cancer in Pap smear images as shown in Fig.\ref{Figure 1}. We first resize and augment the images using a few well-known augmentation techniques that are further described. The input image is flattened into a data array of M patches. When a self-attention operation is carried out for the M elements in a typical Transformer model, the complexity of this operation is $O(M^2)$. However, the proposed model creates a latent array of size N elements, where $N\ll M$, and performs cross-attention and self-attention operation iteratively. A high-dimensional data array is mapped to a smaller dimensional latent bottleneck using a cross-attention block. When the dimensionality is reduced, it is processed through a self-attention based Transformer to create a new lower dimensional latent bottleneck. This process is iteratively repeated to get a final latent space representation of the data.
%First, cross-attention is performed between the latent array and the data array whose complexity is $O(MN)$. Second, self-attention mechanism on the latent array whose complexity is $O(N^2)$.
The proposed method is trained and tested on two benchmark datasets, namely the Sipakmed \cite{plissiti2018sipakmed} and Herlev \cite{jantzen2005pap} Pap smear datasets. Accuracy, precision, F1-Score, and recall are calculated on the test dataset to evaluate performance of the model. The performance of the proposed method is preferable to state-of-the-art methods, according to computational results.
The primary contributions of this paper are listed below:
\begin{enumerate}
  \item The proposed model is based on Transformers and hence requires minimal architectural assumptions about the correlations between the inputs.
  \item The model makes use of an asymmetric attention mechanism to repeatedly consolidate a higher-dimensional input into a lower latent feature space without losing relevant information about the input.
  \item It does away with the quadratic scaling issue of a traditional Transformer and simplifies the depth of the network from the size of the input, enabling us to build extremely deep models.
  \item The proposed model performs better than several state-of-the-art approaches, when evaluated on two standard and readily accessible cervical cancer datasets.
\end{enumerate}
The structure of this paper is as follows: In Section 2, we elaborate on related studies on Pap smear classfication and Transformer architectures. In Section 3, we describe the proposed methodology in detail. In Section 4, results of the proposed method is discussed, including the computational settings and evaluation parameters. In Section 5, we conclude this study and put forward future research prospective.

\section{Prior art}
\subsection{Pap smear classification }
A cervical cancer cell classification technique utilizing convolutional neural networks was proposed by Ghoneim et al \cite{ghoneim2020cervical}. For the feature extraction phase, CNN was used. The input images were then classified into normal and abnormal categories by the extreme learning machine (ELM). The Herlev database was used for the experiments. The proposed CNN-ELM-based system performed the classification of 7 classes with an accuracy of 91.2\%.\\  
Mesquita et al \cite{mesquita2017pap} proposed a cervical cancer detection system that relies on texture information retrieved using randomized neural network signature (RNNS). On the Herlev dataset, the findings showed a classification accuracy of around 87.75\%. The drawbacks include the excessive number of input parameters and rotation sensitivity.\\
In \cite{nguyen2019biomedical}, the cervical cells were classified using feature concatenation and ensemble learning with ResNet152, InceptionV3, and InceptionResNetV2. Performance evaluation was conducted using  Pap smear, 2D Hela dataset, and Hep-2 cell imaging datasets. Each image was scaled to 256 $\times$ 256 pixels. The Herlev dataset had an accuracy score of 93.04\%, and the overall result of ensemble of models is superior to a single model.\\
In \cite{promworn2019comparisons}, a comparative study utilizing deep learning to categorize images of cervical cells was suggested. This work assessed the performance on DenseNet161, ResNet101, DenseNet161, AlexNet, VGG19, and SqueezeNet11 models using the Herlev dataset. DenseNet161 performed the best among the models with accuracy ratings of 68.54\% and 94.38\% for 7-class and 2-class classification, respectively.\\
Srishti et al. \cite{gautam2018cnn} proposed a patch-based CNN classifier to handle the 2-class classification problem. The classifier obtained an F-score of 0.90 on a single cell-level pap smear dataset.\\
In \cite{gautam2018considerations}, a cervical cell analytical method for segmentation, classification, and detection was suggested. The Herlev dataset was classified using transfer learning in this study using AlexNet. They showed that segmentation and classification were not always correlated. Both the 2-class and  7-class classification tasks had accuracy rates of up to 99.3\% and 93.75\%, respectively.\\
A method that integrates machine learning and CNN to categorize images of cervical cells was proposed in \cite{hyeon2017automating}. The features were extracted by the VGG model from the images, and a support vector machine (SVM) classifier was utilized to categorize the data.  A private dataset comprising 71344 images of cervical cells was used in the aforementioned work. The classification outcome had an  F1-Score of  78\%. \\
In \cite{win2020computer}, cervical cancer classification methodology combined SVM, linear discriminant, bagged trees, k-nearest neighbours and boosted trees into an ensemble classifier for classification. The accuracy rate of the Sipakmed dataset for the 2-class and 5-class classifications were 98.27\% and  94.09\% respectively.
\subsection{Effective attention architectures}
ConvNets have turned out to be the dominant group of networks for perceptual tasks since last ten years because of their superior performance and scalability \cite{krizhevsky2017imagenet}. They are able to handle high-resolution images with a minimal number of parameters and computations by sharing weights in 2D convolutions and restricting the computation of each unit to an adjacent 2D neighborhood. However, ConvNets provide a small amount of flexibility when mixing several signals, unlike the Transformers models based on attention \cite{vaswani2017attention}.\\
Although remarkably flexible, Transformers do not scale well with input size, because an equal number of inputs are present in each self-attention layer and are compared to one another at every layer. However, self-attention is increasingly being included into perception, for instance as part of convolutional networks in images \cite{bello2019attention,cordonnier2019relationship,srinivas2021bottleneck} and videos \cite{wang2018non,girdhar2019video}. The input size fed to the Transformer has been reduced using a number of methods, such as convolutional preprocessing \cite{wu2020visual} or subsampling \cite{chen2020generative}, allowing it to be utilized on domains that would otherwise be too huge. Simliar approach has been taken by Vision Transformer (ViT) \cite{dosovitskiy2020image}. It uses a 2D convolutional layer to reduce the input image size by dividing it into fixed size patches, properly embeds each patch before feeding them to a Transformer encoder. The resulting patches are provided with a class token similar to BERT \cite{devlin2018bert}. ViT produces excellent outcomes on ImageNet, but its methodology restricts it to domains that resemble images and have grid-like sample patterns.\\
The Set Transformer \cite{lee2019set} is most closely associated with our work. The Set Transformer projects a high-dimensional input array to a lower-dimensional array via cross-attention, either to ramp up computation inside a module or map the input to a desired output shape.
In a comparable approach without employing cross-attention module, Linformer \cite{wang2020linformer} creates linearly complex self-attention units by mapping key inputs into arrays that are lower-dimensional than the input. The convolutional neural network-inspired Image Transformer \cite{parmar2018image} limits the receptive area of self-attention to adjacent neighborhoods. It makes sure that the likelihood loss is kept within accepted limits when the model is scaled up to handle larger batch sizes. Axial Transformer \cite{ho2019axial} processes large inputs arranged as multidimensional tensors using factorization in a simplified yet efficient approach. It merely applies several attentions, each across a single axis of the input tensor, as opposed to paying attention to the flattened version of the input. Linear Transformer \cite{katharopoulos2020transformers} uses the associative matrix property and a kernel based self-attention architecture to minimize the quadratic scaling issue of self-attention.

%https://towardsdatascience.com/understanding-latent-space-in-machine-learning-de5a7c687d8d
\section{Methodology}
\subsection{Overview}
\begin{figure*}
	\centering  \includegraphics[width=\textwidth,height=3in]{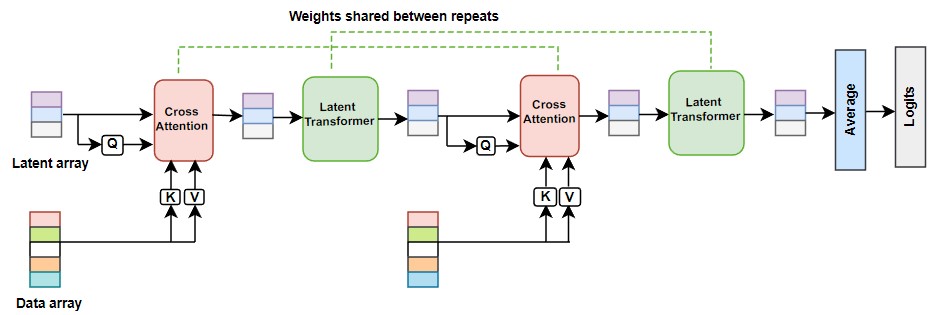}
	\caption{Architectural diagram of the CerviFormer methodology for cervical cancer image classification}
	\label{Figure2}
\end{figure*}

Our architecture is composed of two parts: (1) a cross-attention block that projects a latent array and a data array to a lower dimensional latent bottleneck, and (2) self-attention mechanism based latent Transformer further projects lower dimensional latent bottleneck from the cross-attention block to a new latent feature space as shown in Fig.\ref{Figure2}. The input image size determines the dimension of the data array and is typically large whereas the latent array is a hyperparameter whose dimension is considerably much smaller. In our study, we have used 256 latents. Our model alternately uses the cross-attention module and latent Transformer block. It is equivalent to processing a higher-dimensional input image via a relatively lower-dimensional latent bottleneck prior to feeding it to a Transformer unit, and further utilizing the resultant to query the input data again. The proposed model is similar to a recurrent neural network (RNN), unwrapped in depth by employing an identical input data, instead of time.
We preferably share weights among each iteration of the cross-attention and Transformer block.

\subsection{Cross-attention for reducing quadratic complexity} 
We have built our architecture based on attention because it makes lesser assumptions of the input data size in comparison to convolutional neural networks, hence widely applicable and effective in practice. The primary issue addressed by the proposed architecture is handling very generic and large inputs by scaling attention architectures.\\
Query-key-value (QKV) attention is used to construct the Transformer and cross-attention modules \cite{bahdanau2014neural}. QKV vectors are commonly multi-layer perceptrons (MLPs) neural networks. The input data array is subjected to the QKV attention, which creates three arrays while maintaining the sequence input length (or index dimensionality) M. The quadratic complexity of the QKV attention is the primary obstacle to using Transformers on high-dimensional inputs like images. Since the input dimensionality M of images is very high, it increases the computational cost (e.g, 224 × 224 pixels ImageNet images has M = 50176). Due to this, we have avoided applying directly conventional QKV self-attention to the input data array. Here, we compute cross-attention on the input and latent array before feeding it to the Transformer block. In the QKV self-attention operation, Q $\in$ $\mathbb{R}^{M\times D}$, K $\in$ $\mathbb{R}^{M\times C}$, and V $\in$ $\mathbb{R}^{M\times C}$, where the channel diameters are C and D. The complexity of QKV attention operation is $\mathcal{O}(M^2)$ since it requires two matrix multiplications using matrices of significant dimension M. Therefore, we introduced an asymmetric attention mechanism, where Query (Q) is an embedding sequence from a latent array with dimensionality index N (a hyperparameter such that $N\ll M$), and Key (K) and Value (V) are embedding sequences of the input data array. Cross-attention operation as a result has complexity $\mathcal{O}(MN)$ lesser than the conventional approach. 

\subsection{Latent Transformer for uncoupling depth}
Cross-attention block introduces a bottleneck by taking the input shape to the Q network. By exploiting this bottleneck, deep and expensive Transformers are built in the latent feature space at a minimal cost of $\mathcal{O}(N^2)$. Our proposed method is independent of domain-specific assumptions and relies on linear complex layers, hence can be utilized to build much deeper Transformers than conventional Transformers. The complexity of Transformers structured directly on the input data has complexity $\mathcal{O}(LM^2)$, whereas a Transformer built in latent space has complexity $\mathcal{O}(LN^2)$ ($N \ll M$, and L is the number of layers). When both index dimensionality and the number of layers L are taken into account, the complexity of the architecture becomes $\mathcal{O}(MN + LN^2)$. This enables us to develop very deep networks on large amounts of data. The latent Transformer used in our work utilizes the GPT-2 architecture \cite{radford2019language} which is modeled on the decoder of the Transformer architecture used in \cite{vaswani2017attention}. We have used a value of N=256 in our study, enabling our latent Transformer to have a input size equivalent to models widely used in NLP. A learned position encoding is used to initialise the latent array \cite{gehring2017convolutional}.

\subsection{Repetitive cross-attention and weight sharing}
Our ability to effectively model pixels and create deeper Transformers is made possible by the latent array size, however, severity of the bottleneck may limit our ability to have the network fully extract all of the information from the input data. The model could be constructed with numerous cross-attention layers to lessen this effect. It enables 
the latent array to repeatedly extricate data from the input as required. The model can be tweaked as a result to balance costly, but insightful cross-attends opposed to inexpensive, but potentially repetitive latent self-attends. Thus, sharing weights between the consecutive cross-attention and latent Transformer modules can help improve overall efficiency of the model \cite{lan2019albert,dehghani2018universal}.

\section{Results and Discussion}
\subsection{Datasets}
We have trained and tested the proposed model using the following two publicly available Pap smear datasets.
\subsubsection{Sipakmed Dataset}
\begin{figure*}
	\centering
	  \includegraphics[width=\textwidth,height=2.2in]{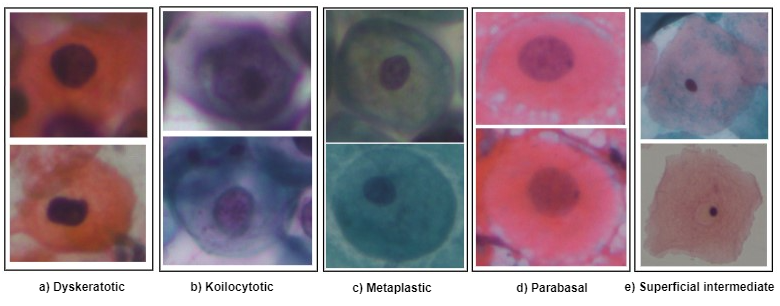}
	\caption{Samples of Sipakmed Dataset:(a) Abnormal (Dyskeratotic, Koilocytotic),(b) Benign (Metaplastic),(c)Normal (Parabasal, Superficial intermediate)}
	\label{FigureSipak}
\end{figure*}

The Sipakmed dataset consists of 4049 distinct images of cervical cells \cite{plissiti2018sipakmed}. Each image has been cropped from the cellular images collected using a charged-coupled device (CCD) camera. The images are further divided into five groups: parabasal, and superficial intermediate, dyskeratotic, koilocytotic, and metaplastic. These five groups can also be categorized into normal, abnormal and benign classes, with the normal  including parabasal and superficial intermediate with 1618 images. Abnormal group consisting of dyskeratotic and koilocytotic categories with 1638 images. Benign category includes the metaplastic with 793 images. Fig. \ref{FigureSipak} displays some samples from the Sipakmed dataset. It can be seen that the cells within the same category in this figure have different colors. Although color information is not significant in cytopathology images, it is crucial for histopathology images. The proposed model performs 3-class classification by taking 80\% of the data in the training set and the remaining 20\% of the data in the testing set. Table \ref{tab1} lists the training and test sets. In order to prevent the overfitting problem, the aforesaid training set of the Sipakmed dataset is augmented by flipping horizontally and randomly zooming by height 0.2 and width 0.2.
\subsubsection{Herlev Dataset}
\begin{figure*}
\centering
\includegraphics[width=\textwidth,height=2.2in]{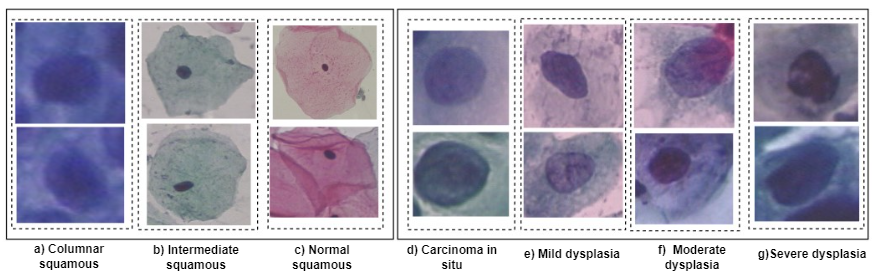}
	\caption{Samples of Herlev Dataset: (i) Normal (a) Columnar squamous (b) Intermediate squamous (c) Normal squamous, (ii) Abnormal (d) Carcinoma in situ (e) Mild dysplasia, (f) Moderate dysplasia (g) Severe dysplasia}
	\label{FigureHerlev}
\end{figure*}

The Herlev dataset \cite{jantzen2005pap} consists of 917 cell images that belong to seven groups: superficial squamous, intermediate squamous, columnar squamous, mild dysplasia, moderate dysplasia, severe dysplasia, and carcinoma in situ. These seven groups can also be divided into normal and abnormal categories \cite{liu2022aspect}, with the former including superficial squamous, intermediate squamous, and columnar squamous with 242 images, and the latter including carcinoma in situ, mild dysplasia, moderate dysplasia, and severe dysplasia with 675 images, as shown in Fig. \ref{FigureHerlev}. The proposed model performs 2-class classification by taking 90\% of the data in the training set and the remaining 10\% of the data in the testing set. 
 Table \ref{tab1} lists the train and test sets. The Herlev dataset is also augmented by flipping horizontally, and random zooming by height 0.2 and width 0.2.  

\begin{table}[h]
%\begin{center}
\begin{minipage}{174pt}
\caption{Number of train and test images}\label{tab1}%
\begin{tabular}{@{}llll@{}}
\toprule
& \hspace{0.8in}Images used\\
& Sipakmed dataset & Herlev dataset\\
\midrule
 Train & 3239 & 825 \\
  Test & 810 & 92\\
\botrule
\end{tabular}
\end{minipage}
%\end{center}
\end{table}

\begin{table}[h]
%\begin{center}
\begin{minipage}{174pt}
  \caption{Description of the Sipakmed dataset}
  \begin{tabular}{@{}llll@{}}
   \toprule
  Category & Class & No. of Images\\
   \midrule
    Normal & Parabasal & 787\\
    Normal & Superficial-Intermediate & 831\\
    Abnormal & Dyskeratotic & 813\\
    Abnormal & Koilocytotic & 825\\
    Benign & Metaplastic & 793\\
    
   \botrule
 \end{tabular}
\end{minipage}
%\end{center}
\end{table}

\begin{table}[h]
%\begin{center}
\begin{minipage}{174pt}
\caption{Description of the Herlev dataset}
\begin{tabular}{@{}llll@{}}
\toprule
    Category & Class & No. of Images\\
   \midrule
    Normal & Superficial squamous & 74 \\
    Normal & Intermediate squamous & 70\\
    Normal & Columnar squamous & 98\\
    Abnormal & Mild dysplasia & 182\\
    Abnormal & Moderate dysplasia & 146\\
    Abnormal & Severe dysplasia & 197\\
    Abnormal & Carcinoma in situ & 150\\
   \botrule
 \end{tabular}
\end{minipage}
%\end{center}
\end{table}

\subsection{Image preprocessing}
The cervical cytopathology cellular images of the Sipakmed dataset are in the BMP format, with sizes ranging from (71 × 59) to (490 × 474) pixels. As a result, for the proposed approach, we have resized the image size to (224 × 224) pixels. Each input image is additionally normalized using Layer normalization. \\
The cervical cytopathology cell images of the Herlev dataset are in the BMP format, with sizes ranging from (77 × 43) to (360 × 577) pixels. As a result, for the proposed approach, we have resized the image size to (72 × 72) pixels. Each input image is additionally normalized using Layer normalization.
\subsection{Computational details}
Our study was carried out on a laptop with Intel(R) Iris(R) Xe graphics, Python 3.7.6, TensorFlow 2.7.0, and Keras 2.7.0. The system also has 16.0 GB of RAM and a 2.40 GHz Intel(R) Core(TM) i5-1135G7 processor.
\subsection{Model parameters}
We have optimized our model using LAMB optimizer, which was developed for training Transformer-dependent models. The proposed model was trained with a 0.001 learning rate, 0.0001 weight decay, and 100 epochs. For our classification purpose, we chose Sparse Categorical Crossentropy as a loss function. We utilized a latent array with the dimensions $256 \times 256$. The ideal batch size for performance is 32. The model parameters are further listed in Table \ref{tab5}.
\begin{table*}[h]
%\begin{center}
%\begin{minipage}{174pt}
\caption{Hyperparameters used in the Proposed model}\label{tab5}%
\begin{tabular}{@{}llll@{}}
\toprule
    Hyperparameters & Sipakmed Data & Herlev Data\\
   \midrule
    Optimizer & LAMB & LAMB \\
    Learning Rate & 0.001 & 0.001\\
    Weight decay & 0.0001 & 0.0001\\
    Loss Function & SparseCategoricalCrossentropy & SparseCategoricalCrossentropy\\
    Batch Size & 32 & 32\\
    Epochs & 100 & 100\\
    Dropout & 0.2 & 0.2\\
    Image size & $224\times224\times3$ & $72\times72\times3$\\
    Patch size & $14\times14\times3$ & $2\times2\times3$\\
    No. of patches(N) & 256 & 1296\\
    Latent dimension(LD) & 256 & 256\\
    Projection dimension(PD) & 256 & 256\\
    Latent array shape($LD\times PD$) & $256\times256$ & $256\times256$\\
    Data array shape($N\times PD$) & $256\times256$ & $1296\times256$\\
\botrule
\end{tabular}
%\end{minipage}
%\end{center}
\end{table*}

\subsection{Evaluation method}
In a study, it is fundamental to select an appropriate assessment method. Particularly, in classification studies, the most commonly implemented and accepted evaluation metrics are precision, F1-Score, recall, and accuracy \cite{sukumar2016computer}. Using the positive-negative binary classification as an explanation, true positives (TP) are defined as the number of positive cases that are correctly predicted. False positives (FP) are the number of negative cases that were incorrectly identified as positive cases. False negatives (FN) are the number of positive cases that were incorrectly identified as negative cases. True negatives (TN) are the number of correctly predicted negative cases. Precision is the fraction of TP among all positive predictions. Recall is the proportion of correctly predicted positive cases to the actual number of positive cases. The F1-Score includes both recall and precision. Accuracy is the proportion of accurate predictions to the total number of cases. The equations of the evaluation parameters are formulated in Table \ref{tab4}.

\begin{table}[h]
%\begin{center}
\begin{minipage}{174pt}
\caption{Evaluation metrics}\label{tab4}%
\begin{tabular}{@{}llll@{}}
\toprule
    Assessment & Formula\\
   \midrule
    Precision(P) & $\frac{TP}{TP+FP}$ \\
    \rule{0pt}{4ex}
    Recall(R) & $\frac{TP}{TP+FN}$\\
    \rule{0pt}{4ex}
    F1-score & $\frac{2*P*R}{P+R}$\\
    \rule{0pt}{4ex}
    Accuracy & $\frac{TP+FN}{TP+TN+FP+FN}$\\
    \botrule
 \end{tabular}
\end{minipage}
%\end{center}
\end{table}

\begin{figure*}
     \centering
     \begin{subfigure}[b]{0.47\textwidth}
         \centering
    \includegraphics[width=\textwidth]{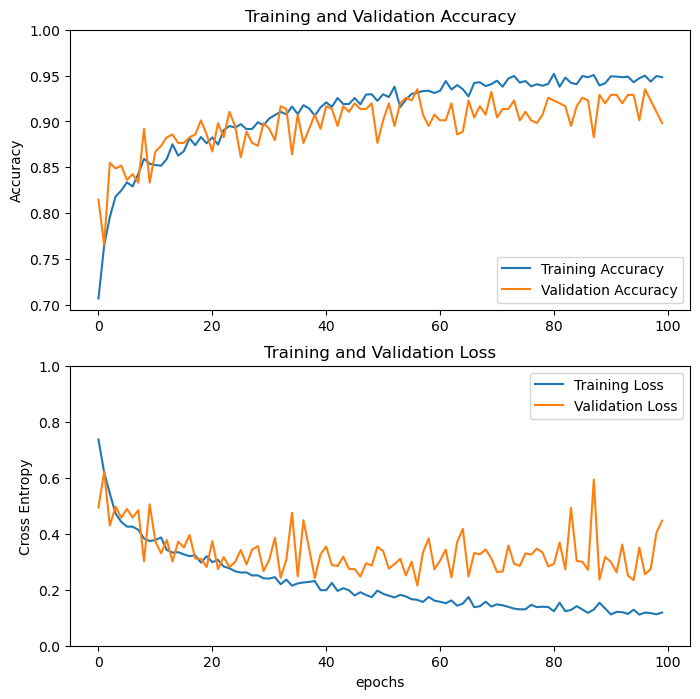}
         \caption{Convergence pattern of training and validation accuracy and sparse categorical crossentropy loss function using proposed model up to epoch 100 in Sipakmed dataset}
         \label{sipak_cm}
     \end{subfigure}
     \hfill
     \begin{subfigure}[b]{0.47\textwidth}
         \centering
         \includegraphics[width=\textwidth]{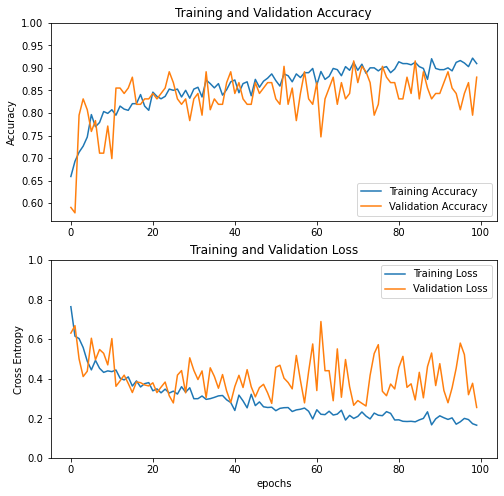}
         \caption{Convergence pattern of training and validation accuracy and sparse categorical crossentropy loss function using proposed model up to epoch 100 in Herlev dataset}
         \label{VIT_PR}
     \end{subfigure}
        \caption{(a) Accuracy vs epoch (Top), Loss vs epoch (Bottom) (b) Accuracy vs epoch (Top), Loss vs epoch (Bottom) }
        \label{FigLoss}
\end{figure*}

\subsection{Results}
\begin{figure*}
     \centering
     \begin{subfigure}[b]{0.47\textwidth}
         \centering
    \includegraphics[width=\textwidth]{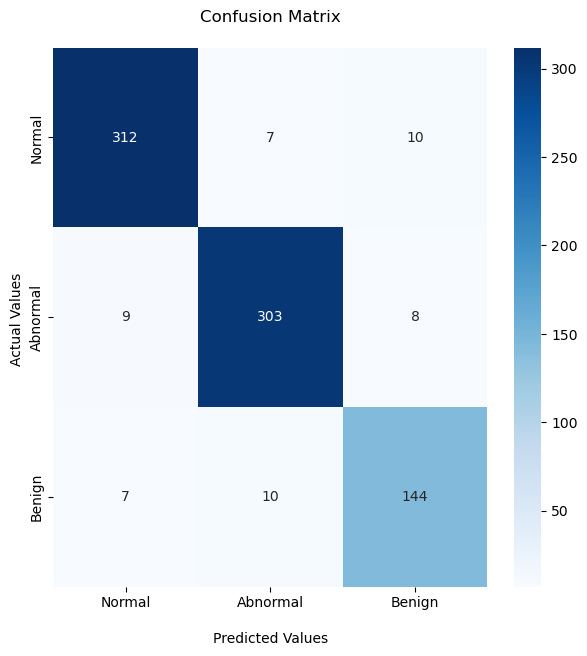}
         \caption{Sipakmed CM}
         \label{sipak_cm}
     \end{subfigure}
     \hfill
     \begin{subfigure}[b]{0.47\textwidth}
         \centering
         \includegraphics[width=\textwidth]{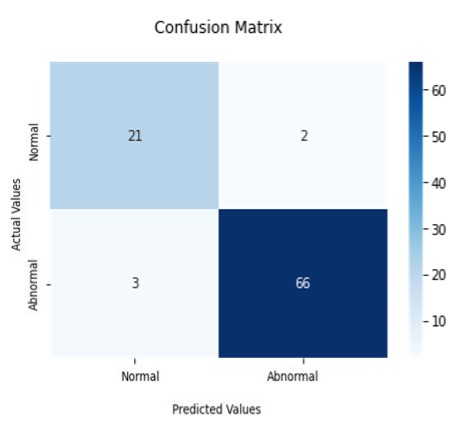}
         \caption{Herlev CM}
         \label{VIT_PR}
     \end{subfigure}
        \caption{(a) Confusion matrix (CM) of the Sipakmed dataset. (b) Confusion matrix (CM) of Herlev dataset }
        \label{FigCM}
\end{figure*}

\textbf{Sipakmed data }Fig. \ref{FigLoss}(a) displays the graphical representation of the convergence pattern of training and validation accuracy of the proposed model together with loss up till 100 epochs. This ensures the robustness of the model and it does not undergo overfitting. The overall 3-class classification accuracy score is 93.70\% based on the confusion matrix of Sipakmed data displayed in Fig. \ref{sipak_cm}. We observe that out of 810 testing cases, 759 are accurately identified and classified into 3 classes. Then, as shown in Table \ref{tab6}, we compute the precision, recall rate, and F1-score of the proposed model based on asymmetric attention. The computed Cohen's kappa coefficient is 0.90, which further supports the validity of the proposed approach. The automated method produces 92.99\% sensitivity and 96.81\% specificity, which can be used to further assess how well our proposed method performs in predicting cervical cancer cases using pap smear images.\\
\textbf{Herlev data }Fig. \ref{FigLoss}(b) displays the graphical representation of the convergence pattern of training and validation accuracy of the proposed model together with loss up till 100 epochs. The overall 2-class classification accuracy score is 94.57\% based on the confusion matrix of Herlev data displayed in Fig. \ref{FigCM}(b). It shows that there are 21 true negatives (TN), 2 false positives (FP), 3 false negatives (FN) and 66 true positives (TP). We observe that out of 92 testing cases, 87 are accurately identified and classified into 2 classes. Then, as shown in Table \ref{tab7}, we compute the precision, recall rate, and F1-score of the proposed model based on asymmetric attention. The computed Cohen's kappa coefficient is 0.85, which further supports the validity of the proposed approach. The automated method produces 87.5 \% sensitivity and 97.06 \% specificity, which can be used to further assess how well our proposed method performs in predicting cervical cancer cases using pap smear images as shown in Table \ref{tab8}.\\
Additionally, the methodologies described in this work have been contrasted with some of the earlier suggested techniques for classifying pap smear cervical cell images. Table \ref{compare} compares the experimental outcomes from the current work with those from previous methodologies. As shown in Table \ref{compare}, the proposed method outperforms existing state-of-the-art methods. As a result, the proposed method performs well on a range of performance indicators, demonstrating their efficacy.

\begin{table}[h]
\begin{center}
%\begin{minipage}{174pt}
\caption{Evaluation metrics of Sipakmed dataset}\label{tab6}%
\begin{tabular}{@{}lllll@{}}
\toprule
Class & Precision & Recall & F1-Score\\
\midrule
Normal & 0.95 & 0.95 & 0.95\\
Abnormal & 0.95 & 0.95 & 0.95\\
Benign & 0.89 & 0.89 & 0.89 \\
\botrule
\end{tabular}
%\end{minipage}
\end{center}
\end{table}

\begin{table}[h]
\begin{center}
%\begin{minipage}{174pt}
  \caption{Evaluation metrics of Herlev dataset}\label{tab7}%
  \begin{tabular}{@{}lllll@{}}
\toprule
    Class & Precision & Recall & F1-Score\\
   \midrule
    Normal & 0.88 & 0.91 & 0.89\\
    Abnormal & 0.97 & 0.96 & 0.96\\
   \botrule
\end{tabular}
%\end{minipage}
\end{center}
\end{table}

\begin{table}[h]
\begin{center}
%\begin{minipage}{174pt}
  \caption{Performance metrics of CerviFormer methodology}\label{tab8}%
  \begin{tabular}{@{}lllll@{}}
\toprule
    Dataset & Accuracy & Specificity & Sensitivity\\
   \midrule
    Sipakmed & 93.70\% & 96.81\% & 92.99\%\\
    Herlev & 94.57\% & 97.06\% & 87.50\%\\
   \botrule
\end{tabular}
%\end{minipage}
\end{center}
\end{table}

%\begin{table}[h]
%\begin{center}
%\begin{minipage}{174pt}
%\caption{State-of-the-art comparison on Sipakmed dataset.}\label{tab4}%
%\begin{tabular}{@{}llll@{}}
%\toprule
    %Approach & Accuracy\\
   %\midrule
    %\cite{haryanto2020utilization} & $87.32 \%$ \\
    %\rule{0pt}{4ex}
    %\cite{win2020computer} & $xx.09\%$ \\
    %\rule{0pt}{4ex}
    %\cite{nguyen2019biomedical} & $xx.04\%$\\
    %\rule{0pt}{4ex}
    %\cite{promworn2019comparisons} & $xx.38\%$\\
    %\rule{0pt}{4ex}
    %\textbf{Proposed work} & $93.70\% $\\
    %\botrule
 %\end{tabular}
%\end{minipage}
%\end{center}
%\end{table}
%94.09%

\begin{table}[h]
%\begin{center}
\begin{minipage}{174pt}
\caption{The comparison with state-of-the-art methods}\label{compare}%
\begin{tabular}{@{}llll@{}}
\toprule
    Approach & Accuracy\\
   \midrule
   \cite{haryanto2020utilization} & $87.32 \%$ \\ %sipak
    \rule{0pt}{4ex}
    \cite{mesquita2017pap} & $87.75\%$ \\
    \rule{0pt}{4ex}
    \cite{ghoneim2020cervical} & $91.20\%$ \\
    \rule{0pt}{4ex}
    \cite{nguyen2019biomedical} & $93.04\%$\\
    \rule{0pt}{4ex}
    \cite{promworn2019comparisons} & $94.38\%$\\
    \rule{0pt}{4ex}
    \textbf{Proposed work(Sipakmed Dataset)} & $93.70\% $\\
    \rule{0pt}{4ex}
    \textbf{Proposed work(Herlev Dataset)} & $94.57\%$\\
    \botrule
 \end{tabular}
\end{minipage}
%\end{center}
\end{table}

\subsection{Discussion}
As a result of our research, it can be said that computerised binary/multi-class prediction of cervical cancer can pave the way for a decision-support system that can help pathologists determine the severity of the condition and the best course of treatment. The experimental outcomes further show that the proposed model outperforms other alternative classifier models in terms of classification accuracy, precision, and recall for the 3/2 target classes. In order to emphasize the proposed method, we have highlighted the key results below based on our analysis:
\begin{enumerate}
    \item We have proposed CerviFormer, a methodology that can accept very high dimensional inputs. This creates unique opportunities for generic Transformer-related models that can work with very few input assumptions, and accommodate variable sensor configurations while facilitating the merging of information at all levels.
    \item Our key focus is the formation of attention bottleneck by introducing a finite number of latent modules across which the inputs must transit. As a result, the quadratic scaling issue of a standard Transformer is resolved and it further permits the formation of extremely deep models by breaking the relationship between the depth of the network and the size of the input.
    \item The suggested model iteratively focuses its limited resources on the most significant inputs. However, many modalities require spatial or temporal information, and it is often necessary to identify input among various modalities in multimodal contexts. The absence of explicit structures in our framework have been compensated by creating modality and position-specific features for every input element such as pixel.
    
    % This technique of tagging input units is quite comparable to the labeled lined approach used in the construction of cross-sensory and topographic maps in biological neural networks \cite{kandel2012principles}.
    
\item This approach enriches the framework since it can classify entire slides of pap smear images without depending on segmentation methods.
\end{enumerate}

\section{Conclusion}
The traditional Pap smear procedure is still the most affordable and widely utilized approach, both for patients diagnosed and for periodic early diagnosis camps hosted by governmental and non-governmental groups. Therefore, the development of an effective and reliable automated Pap smear screening system would be highly beneficial in the healthcare domain. However, this method is inefficient in tackling confounding factors such as menstrual cycle and age which can also affect the output of the system. But future research will take these confounding factors into account as newly added features. Additionally, we will emphasize increasing the size of the data set so that overall accuracy can be kept consistent when working with images that have different degrees of dysplasia. Future studies will also test the method on ThinPrep images.\\
In this work, the integrated cross-attention and latent Transformer model, known as CerviFormer, is implemented for the classification of Pap smear images. It performs with an accuracy of 93.70\% (3 classes) using the Sipakmed database and 94.57\% (2 classes) using the Herlev database. All of the results indicate that the proposed methodology is very promising for assessing the degree of dysplasia present in cervical lesions using Pap smear images. This form of automated method boosts efficiency, eliminates observer bias, and reduces the amount of time needed for manual observations.\\

\bibliographystyle{sn-mathphys}
\bibliography{sn-bibliography}% common bib file


\section*{Acknowledgments}
%\bmhead{Acknowledgments}

% This research received no specific support from governmental, private, or non-profit funding agencies. The content and writing of the paper are solely the responsibility of the authors. 
Bhaswati Singha Deo is thankful to IIT Kanpur for the Institute fellowship. The author MPal wishes to thank ABB Ability Innovation Centre, India for their support in this work.

\section*{Data availability}
%\bmhead{Data availability}
The obtained datasets for our analysis are from free public database. \\ Sipakmed data: \url{https://www.cs.uoi.gr/~marina/sipakmed.html}.\\
Herlev data:  \url{https://www.kaggle.com/datasets/yuvrajsinhachowdhury/herlev-dataset}.

\section*{Statements and Declarations}

\bmhead{Funding}
This research received no specific support from governmental, private, or non-profit funding agencies.

\bmhead{Conflict of interests}
The authors declare that they have no known competing financial interests or personal relationships that could have appeared to influence the work reported in this paper.

\bmhead{Author contributions}
Bhaswati Singha Deo: Developed the model from the
concept, Developed the code, Generated the results, Wrote the manuscript. Mayukha Pal: Conceived the idea and conceptualized it, Developed the methodology from the concept, Reviewed the manuscript and mentored the work.
Prasanta K. Panigrahi: Reviewed the manuscript and mentored the work. Asima Pradhan: Developed the analysis methodology, Reviewed the manuscript and mentored the work.

\bmhead{Ethics approval}
Datasets are publicly available for research purpose, hence not applicable for this paper.

\bmhead{Consent to participate}
Datasets are publicly available for research purpose, hence not applicable for this paper.

\bmhead{Consent to publish}
Datasets are publicly available for research purpose, hence not applicable for this paper.



\end{document}